# The Impact of LLM-Generated Reviews on Recommender Systems: Textual Shifts, Performance Effects, and Strategic Platform Control

**Itzhak Ziv, Moshe Unger and Hilah Geva, Tel-Aviv University**

## Abstract

The rise of generative AI technologies is reshaping content-based recommender systems (RSes), which increasingly encounter AI-generated content alongside human-authored content. This study examines how the introduction of AI-generated reviews influences RS performance and business outcomes. We analyze two distinct pathways through which AI content can enter RSes: *user-centric*, in which individuals use AI tools to refine their reviews, and *platform-centric*, in which platforms generate synthetic reviews directly from structured metadata. Using a large-scale dataset of hotel reviews from TripAdvisor, we generate synthetic reviews using LLMs and evaluate their impact across the training and deployment phases of RSes. We find that AI-generated reviews differ systematically from human-authored reviews across multiple textual dimensions. Although both user- and platform-centric AI reviews enhance RS performance relative to models without textual data, models trained on human reviews consistently achieve superior performance, underscoring the quality of authentic human data. Human-trained models generalize robustly to AI content, whereas AI-trained models underperform on both content types. Furthermore, tone-based framing strategies (encouraging, constructive, or critical) substantially enhance platform-generated review effectiveness. Our findings highlight the strategic importance of platform control in governing the generation and integration of AI-generated reviews, ensuring that synthetic content complements recommendation robustness and sustainable business value.

**Keywords:** Recommender systems, LLMs, Generative AI, content-based RS, content quality, collaborative filtering, reviews, cold-start, sparsity

# 1    INTRODUCTION

The rapid rise of generative artificial intelligence (GenAI) technologies, particularly Large Language Models (LLMs), is reshaping modern information systems (Chatterji et al., 2025), including content-based recommender systems (RSes) (Wu et al., 2024; Zhao et al., 2024). These systems, which are instrumental to review-based platforms such as TripAdvisor, Yelp, and Amazon, leverage user-written reviews to suggest relevant items and improve personalization. Traditionally built on human-generated content, RSes are now increasingly likely to encounter synthetic, AI-generated input, which, as we elaborate below, can be introduced both through individual users (e.g., a consumer who uses AI to improve the writing style of a review she has written) and through the platforms themselves (e.g., leveraging generative models to generate synthetic reviews to enhance personalization and align with human-like content).

The shift from human-generated to AI-generated review content challenges long-standing assumptions on which RS algorithms are based, regarding the authenticity, credibility, and diversity of reviews, and more broadly, the value of review input as an encapsulation of the "wisdom of the crowd" (Santos and Antonio, 2025; Wen and Laporte, 2025). In concrete terms, these concerns manifest as information quality challenges with regard to RS training and deployment. Specifically, there may be systematic differences between AI-generated and human-generated reviews along various dimensions, including content, style, tone, quality, and consistency (Santos and Antonio, 2025; Shan et al., 2025; Yu et al., 2023; Wen and Laporte, 2025), which could diminish RSes' capacity to leverage reviews to learn about users' actual preferences. Such differences may, in turn, degrade model performance, observable in RSes' capacity to predict users' ratings and rankings of specific products, as well as in business-facing measures such as the quality of recommended items, popularity exposure, perceived helpfulness of surfaced content, and the diversity of recommendations. Accordingly, the rise in AI-generated review content has significant implications for the future design and evaluation of RSes. However, to our knowledge, researchers have yet to examine how the introduction of AI-generated data, when originating from users or platforms, into the recommendation pipeline influences RS performance. Herein, we begin to address this critical gap.

In this study, we assume that GenAI content can enter an RS through two distinct pathways, each with different implications for RS performance, platform strategy and system design. The first, which we term user-centric, arises when individual users employ tools such as ChatGPT or Gemini to refine, enhance, or even generate their own reviews. While this assistance may improve clarity and tone, it can also introduce systematic textual patterns, such as reduced lexical diversity or homogenized sentiment, that diverge from natural human expression. Because RSes were originally optimized to learn from authentic user-generated language, these distributional shifts may subtly degrade model performance or bias outcomes in ways that are not immediately visible to platforms or end users.

The second pathway, which we term platform-centric, arises when the platform itself generates or supplements review content. This pathway differs fundamentally from the user-centric pathway because it grants platforms broad control over the creation, framing, and deployment of textual signals. In some cases, such platform-generated content is visible to users: For example, platforms may offer AI-drafted reviews to users who are reluctant to write, allowing individuals to post these drafts at their discretion. This semi-automated form of assistance can boost review volume while still preserving user agency. Other forms of platform-centric AI content generation take place "behind the scenes", during the algorithm training phase. Specifically, platforms can generate entirely synthetic reviews from structured interaction data (e.g., ratings, attributes, or item metadata) and inject them directly into the training pipeline of recommendation models. In this case, the generated content is not visible to end users but serves as an additional textual signal that shapes how the RS learns user preferences and item relationships. Platforms may pursue this pathway because it addresses "cold-start" conditions, i.e., situations in which insufficient textual review content is available to train the RS. In such cases, even synthetic review content can provide informative textual patterns and tonal variations that help the model learn more effectively.

In addition to alleviating challenges of data sparsity on both the training and deployment ends, platform-centric AI content generation introduces a powerful new layer of platform control, where synthetic text, including content and tone, can be strategically designed to align with platform objectives (Yu et al., 2023). At the same time, such interventions risk undermining the very system they are designed to support, by altering the statistical properties of the content in ways that degrade model generalization, exacerbate popularity bias, or promote specific items in unintended ways (Zhou et al., 2025). They can also create mismatches between training and deployment distributions, for example, when models trained on human-authored reviews are increasingly exposed to GenAI-crafted content at deployment time (or vice versa), leading to unpredictable changes in accuracy and business outcomes.

Thus, both user-centric and platform-centric GenAI review content can reshape either RS training data or the real-world platform data on which RSes are deployed in practice, potentially affecting the performance of RSes that typically operate on human-generated reviews. Our main goal is to assess this influence. Accordingly, a necessary first step is to gain insight into the nature of the discrepancies between AI- and human-generated reviews. This leads to our first research question:

***RQ1: How do AI-generated reviews systematically differ from traditional human-written reviews, and how do these differences manifest across user- and platform-generated contexts?***

Building on these differences, we then proceed to address how various forms of GenAI reviews affect the performance of RS algorithms, in terms of their prediction accuracy as well as their contribution to desired platform-level business outcomes. First, we focus on user-centric scenarios, where individuals rely on AI tools to craft or enhance their reviews. In these scenarios, the resulting text resembles authentic feedback while reflecting generic language-model patterns, yet it is explicitly designed to align with users' ratings across multiple categories, preserving the evaluative intent of the original input. Such content could diminish the capacity of RSes to learn user preferences from the reviews that they have "authored" (e.g., to conclude which star rating a user will assign to a particular product, given the text of the user's prior reviews). Accordingly, we ask:

***RQ2: In user-centric scenarios, how does the use of AI-generated reviews affect recommendation performance and business outcomes?***

We then turn to platform-centric content, where the challenges may be amplified even further, given that such content introduces a level of editorial control (e.g., over tone and textual content) that is unlikely to arise organically in reviews generated independently by individual users. Here, we focus on the generation and integration of fully synthetic reviews into the RS pipelines. We approach this analysis in three parts. First, similarly to the user-centric scenario, we seek to understand how the replacement of human-authored reviews with (neutral-toned) platform-centric GenAI reviews influences recommendation performance and business outcomes. Next, we examine cross-training settings in which the training dataset and the deployment dataset are of different origins (platform-centric GenAI reviews vs. human-generated reviews, respectively, or vice versa). For example, such a mismatch could arise when a platform adopts a model pre-trained on human-authored reviews but deploys it on synthetic, platform-generated reviews. These types of mismatches may cause models to underperform, misgeneralize, or exhibit distributional biases that would appear in the deployment phase. Finally, we explore scenarios in which the platform not only generates (neutral-toned) synthetic reviews but also exploits the opportunity to fine-tune the tone of these reviews. Here, by modeling the platform as an active generator, curator, and strategist, we aim to capture how GenAI content reshapes the dynamics of RSes. Formally, our final research question is:

***RQ3: In platform-centric scenarios:***

**(a) *How does the use of AI-generated reviews affect recommendation performance and business outcomes?***

**(b) *How do cross-training scenarios, where models are trained on one review source (human or AI) but tested on another, affect recommendation performance and business outcomes?***

**(c) *What are the effects on RS performance and business outcomes when platforms strategically manipulate review tone (e.g., to elicit encouraging vs. critical) in platform-generated reviews?***

To study these questions, we focus on the travel domain, where online reviews strongly influence consumer decisions and platform credibility. Specifically, we collected a large-scale dataset of hotel reviews from TripAdvisor as our corpus of human-authored reviews. We then used a state-of-the-art LLM to produce AI-generated reviews under controlled prompting conditions that reflected both our user-centric and platform-centric scenarios, including platform-centric scenarios in which reviews were designed to adopt a specific tone. In all cases, the generated reviews were explicitly aligned with users' multi-aspect rating profiles to ensure consistency between textual and numerical evaluations. We then proceeded to explore RQ1, analyzing human-generated, user-centric-AI-generated, and platform-centric-AI-generated reviews along multiple dimensions to identify systematic differences between them. We then used our various review corpora to explore RQ2 and RQ3. In the latter analyses, we implemented a traditional neural content-based RS and evaluated its performance under different conditions, using both standard accuracy metrics (e.g., RMSE and MAE), ranking metrics (e.g., NDCG and MRR), and business-oriented measures. This design enables us to isolate the effects of content origin, tone framing, and cross-training strategies on RS performance and business outcomes.

We find that AI-generated reviews differ systematically from human-authored reviews across multiple textual dimensions (RQ1). Despite these differences, both user-centric and platform-centric AI-generated reviews significantly improve recommendation performance and business outcomes compared to models trained without textual data. However, models trained on human-authored reviews consistently outperform those trained on AI-generated reviews across rating accuracy, ranking quality, and strategic business indicators such as hotel quality and popularity exposure (RQ2, RQ3.a). Importantly, we demonstrate that training on human-authored (vs. AI-generated) reviews provides greater robustness: models trained on human reviews generalize effectively even when evaluated on AI-generated content, whereas models trained on AI-generated reviews perform less well on both AI and human content, though notably, even cross-training scenarios with AI-generated reviews still outperform the no-review baseline (RQ3.b). Finally, we show that tone-based framing strategies, using encouraging, constructive, or critical tones, can meaningfully enhance the effectiveness of platform-generated reviews, narrowing the performance gap in comparison to human-authored content (RQ3.c). Together, these findings reveal that human-authored reviews remain the gold standard for recommendation quality, but AI-generated reviews offer substantial improvements over no textual data and represent a valuable tool for platforms facing content scarcity. Moreover, simple shifts in reviews' tones show promise for further enhancing the effectiveness of RSes.

Our study makes several interrelated contributions to research on GenAI and RSes. First, from a theoretical perspective, we extend the literature on content-based RSes by examining how GenAI content integrates into and impacts RS pipelines. Specifically, we relax a foundational assumption in prior research, that the textual inputs used for personalization are human-authored. Existing RS studies have typically treated reviews as faithful, heterogeneous reflections of user preferences (Liu et al., 2020; Cai et al., 2022; Hasan et al., 2025), without considering how machine-generated texts alter the informational properties on which recommendation learning depends. We introduce GenAI content origin as a new dimension of review data, recognizing that machine-generated reviews constitute a distinct source of textual input within RSes. Incorporating this distinction extends RS theory to include hybrid learning environments in which human- and machine-generated reviews coexist, linking IS research on generative AI's linguistic patterns with established findings on how textual inputs shape RS performance.

Second, we develop a conceptual framework that introduces two distinct pathways through which GenAI content integrates into RS pipelines: a ***user-centric*** pathway, in which individuals employ LLMs to refine or generate their own reviews, and a ***platform-centric*** pathway, in which platforms generate synthetic reviews directly from structured data and metadata available to them. These pathways differ in the degree of user versus platform control, the manner in which synthetic content is created, and the tone and style of the generated text. By distinguishing between them, the framework demonstrates that the effects of GenAI content on RS performance depend not only on its linguistic characteristics, but also on its origin (user vs. platform) and on how it is incorporated into the RS pipeline (i.e., during training or deployment phase).

Further, we provide an empirical contribution by testing this conceptual framework using a large-scale dataset of hotel reviews from TripAdvisor to evaluate how GenAI reviews affect

RSes' performance. We are among the first to systematically examine how AI-generated reviews, designed to align with users' ratings, influence both recommendation performance metrics (e.g., RMSE, NDCG) and business-relevant outcomes such as helpfulness and popularity.

Finally, from a managerial perspective, we provide actionable guidance for platform owners. Our findings suggest that AI-generated reviews can enhance both recommendation quality and business performance, yet realizing these benefits requires deliberate platform control. Such reviews should complement rather than replace human-authored content, as RS models trained on authentic reviews remain more robust and generalizable. Platforms should implement mechanisms, particularly during the training phase, to detect and align AI-generated content, ensuring that synthetic reviews reinforce rather than distort preference learning. Moreover, tone-aware review generation, achieved by strategically guiding GenAI outputs toward encouraging, constructive, or critical expressions, can further improve recommendation quality and narrow the gap between synthetic and human-authored reviews. Together, these insights position GenAI content integration as a strategic platform decision: determining not only how synthetic reviews are produced, but also how human and AI-generated data are balanced within the RS pipeline to ensure robust performance and sustainable business outcomes.

The paper proceeds as follows. We first review relevant literature on GenAI in IS and on content-based RSes. We then present our model architecture, describe the data, and detail our methodology for generating and evaluating AI-generated and human-authored reviews. Finally, we present the empirical results, followed by a discussion of managerial implications, limitations, and future research.

## 2    LITERATURE

This section reviews two streams of research relevant to our study. The first examines the characteristics and implications of AI-generated content in IS, and the second focuses on content-based RSes and the integration of LLM-generated reviews into RS pipelines.

## 2.1   GenAI Content in IS

Recent IS research has begun to characterize the structural, lexical, and emotional differences between human-authored and LLM-generated content. AI-generated texts have been shown to diverge significantly from human-authored ones, particularly in their failure to emulate natural communication patterns, which enables accurate authorship detection (Sardinha, 2024). Human-written news articles were found to display greater variation in sentence length, richer vocabulary, shorter constituents, and more optimized dependency distances than AI-generated articles, and to express stronger negative emotions and less joy (Muñoz et al., 2024). AI-generated marketing narratives were reported to convey more positive sentiment than human-generated narratives and to influence purchase intent, yet lacked the lexical diversity and embodied cognition present in human-authored content (Wen and Laporte, 2025). AI-generated reviews were observed to receive higher ratings, and to exhibit greater inconsistency between sentiment and rating (Shan et al., 2025). AI-paraphrased hotel reviews were found to be more textually similar to each other than to their original human counterparts, suggesting a homogenizing effect (Xylogiannopoulos et al., 2024).

Collectively, these studies demonstrate that AI-generated content is not simply a replacement for human-authored content: it differs in systematic ways that may affect how it is processed by both users and algorithms. Yet most IS work to date has focused on characterizing these differences and the ways in which users respond to them, leaving open the question of how these differences propagate through algorithmic pipelines, particularly in RSes. Moreover, research has paid little attention to platform-level strategies, such as framing, tone control, or selective generation, that could amplify or mitigate these effects. Our study addresses this gap by systematically comparing human-authored and AI-generated reviews in both user-centric and platform-centric contexts (RQ1), and by evaluating the effects of these different types of reviews on RS performance and business metrics (RQ2 and RQ3).

## 2.2 Content-Based RSes and the Integration of AI-Generated Content

Traditional RSes are often based on collaborative filtering (CF) methods, including matrix factorization (MF) and its deep learning extension, neural collaborative filtering (NCF); these models rely mainly on user–item interactions (He et al., 2017). While effective, these models capture preferences at a coarse level of IDs and ratings, and therefore overlook the more nuanced signals users embed in their written evaluations. This limitation has motivated content-based and hybrid approaches that incorporate human-authored reviews (Zheng et al., 2017; Liu et al., 2020; Cai et al., 2022; Chen and Chen, 2014; Hasan et al., 2025). While early studies combined matrix factorization with review-derived topic or word embeddings, later efforts employed deep neural encoders and hybrid fusion frameworks. Collectively, these works highlight the added value of textual semantics for capturing fine-grained user preferences. These textual semantics in the form of embeddings allow RSes to distinguish why two users may rate an item similarly but for different reasons, to provide more interpretable recommendations, and to offer robustness under sparsity conditions where ratings alone are noisy or insufficient. Empirical work demonstrates that textual reviews improve predictive accuracy by providing complementary signals to ratings and by alleviating sparsity problems (Zheng et al., 2017; Liu et al., 2020; Cai et al., 2022; Fu et al., 2019; Li et al., 2019; Chen and Chen, 2014). Moreover, IS and management studies further confirm that textual signals in reviews influence not only recommendation accuracy but also business outcomes such as sales lift and purchase intent (Yu et al., 2023). Other research shows that the ways in which platforms frame recommendations, such as using personalized textual suggestions instead of aggregate ratings, can also shape users' preferences and engagement (Adomavicius et al., 2022).

Building on this foundation, recent work highlights a broader transition toward integrating LLMs into RSes. Rather than relying solely on structured user–item data, these models leverage LLMs' capacity to extract and reason over rich textual signals, including human-authored user reviews, descriptions, and other unstructured content. LLMs can serve as semantic encoders, transforming human-authored review text into deeper, context-aware representations of preferences, or as generative components that produce synthetic text, such as explanations, summaries, or auxiliary textual features, to enrich recommendations. Recent surveys outline this emerging paradigm, distinguishing between LLM-based recommendation pipelines that use pre-trained models for understanding (encoding) these textual features versus those that employ them for generation or interaction (Zhao et al., 2024; Wu et al., 2024). Encoding-oriented approaches show that enriching point-of-interest recommendations with LLM-based embeddings improves both accuracy and diversity, and that combining reviews with item content helps mitigate cold-start issues in cross-domain settings (Wang et al., 2025; Fu et al., 2019; Hasan et al., 2025). Generative-oriented models, in turn, demonstrate that augmenting user histories with synthetic or imagined items derived from structured metadata can further improve profiling and alleviate sparsity challenges (Liang et al., 2025). Our study complements and adds to this stream of literature by focusing on the effects of generated textual content on RS, rather than generated items. Specifically, we introduce two complementary types of LLM-generated reviews: (1) user-centric reviews, in which human-authored text is refined using LLMs, and (2) platform-centric reviews, in which entirely synthetic content is generated from the multi-aspect structure of user ratings. We systematically evaluate how these generated reviews influence existing RSes across rating, ranking, and business-level metrics, and further analyze how different review tones affect recommendation performance.

AI-generated reviews can provide some of the same semantic signals as human-authored ones, helping to mitigate cold-start challenges, yet they are not equivalent. As discussed earlier, differences between synthetic and human-generated text in tone, style, and lexical diversity have been well documented (Sardinha, 2024; Muñoz et al., 2024; Wen and Laporte, 2025; Shan et al., 2025; Kovács, 2024; Xylogiannopoulos et al., 2024), and these differences can influence how RSes learn user preferences. Although such properties of synthetic reviews may make reviews appear clearer or even more useful, they risk homogenizing the input signals available to RSes. Recent work further suggests that when AI-generated content becomes predominant in the recommendation ecosystem, systems may increasingly favor such content and narrow the diversity of sources, creating a feedback-loop bias (Zhou et al., 2025). While Zhou et al. (2025) analyze the systemic escalation of source bias when AI-refined reviews and authentic human content coexist in sequential RSes, our focus is fundamentally different. We specifically seek to understand the core differences between AI-generated and human-authored reviews and to assess their downstream impact on recommender behavior within CF systems, which remain the most established and widely validated backbone of modern RSes. Furthermore, while Zhou et al. (2025) considered only AI-refined reviews, that is, a simpler version of our user-centric scenario without incorporating additional item metadata, we extend this analysis by examining both user-centric reviews, derived from refined human-authored text aligned with user ratings, and platform-centric reviews, generated directly from structured rating information and other metadata. This dual design moves beyond diagnosing feedback-loop bias and limited refinement settings to examine how the design and origin of generative content, including content synthesized solely from rating and other metadata, shape learning dynamics, personalization, and business-level performance in CF-based RSes.

Relatedly, IS research has underscored the importance of emotional and tonal signals in reviews. Discrete emotions expressed in textual feedback predict key outcomes such as perceived helpfulness and recommendation accuracy more effectively than coarse sentiment polarity (Yu et al., 2023), and recent models explicitly embed emotion and sentiment conditioning into recommendation architectures (Kim and Hong, 2024; Abakarim et al., 2025). Together, this evidence shows that both the content and the style of reviews shape how RSes learn from textual input and influence user engagement. At the same time, prior studies have almost exclusively trained and evaluated their models on human-authored reviews, implicitly assuming that the data used for training and deployment originate from the same source. This design choice largely reflected data availability rather than theoretical consideration, leaving unexamined what happens when RSes trained on human reviews encounter AI-generated content or vice versa. Such cross-training conditions are increasingly likely in hybrid recommendation environments, yet their implications for algorithmic robustness and business outcomes remain unexplored. Accordingly, our study addresses this gap by systematically varying the tone of AI-generated reviews and by examining cross-origin scenarios that test how models trained on one source of text generalize to another (RQ3 b,c).

## 3   MODEL

In this study, we investigate how reviews produced or adjusted by GenAI impact the performance of RSes compared to reviews authored by humans. To this end, we extend the standard Neural Collaborative Filtering (NCF) framework (He et al., 2017) by incorporating either human or GenAI textual reviews alongside user and item identifiers, as illustrated in Figure 1. A critical feature of the design is that review inputs are restricted to historical content available prior to the rating event, ensuring that no future information is leaked into the prediction process. We focus on NCF because it represents a strong, widely adopted baseline for modeling user–item interactions and has been extended in numerous recent studies, including semantic-enhanced variants (Do and Nguyen, 2022; Wu et al., 2022). Moreover, by integrating textual embeddings directly with CF-based user–item signals, this framework enables a robust assessment of how GenAI- versus human-generated reviews influence RS performance.

The NCF model is designed for the rating prediction task and estimates user-item interactions by learning latent embeddings of user and item identifiers, aiming to predict the explicit rating that a given user $u$ would assign to an item $i$. Our extension enriches this baseline by incorporating semantic information extracted from recent user and item textual reviews, as illustrated in Figure 1. At a high level, the extended NCF architecture consists of three main components:

(i) **Identifier embeddings:** As shown in the first input layer of Figure 1, user and item IDs are represented as one-hot vectors and mapped into dense embeddings via trainable embedding layers, similarly to (He et al., 2017). These embeddings, denoted $\mathbf{e}_u$ and $\mathbf{e}_i$, capture CF signals.

(ii) **Review embeddings:** In addition to identifiers, the model incorporates textual reviews from two independent sources: the $k$ most recent reviews written by the user and the $k$ most recent reviews associated with the item. Each review is encoded into a dense $d$-dimensional vector using Sentence-BERT. As shown in the middle part of Figure 1, the concatenated review embeddings are passed through separate multi-layer perceptrons (MLPs), referred to as Learning

Layers in the figure, to denoise, reduce dimensionality, and extract the most informative semantic signals. The resulting vectors are denoted $\widetilde{\boldsymbol{h}}_u$ for user reviews and $\tilde{\boldsymbol{z}}_i$ for item reviews.

(iii) **Prediction network:** Finally, as shown at the top of Figure 1, the identifier embeddings $(\mathbf{e}_u, \mathbf{e}_i)$ and compressed review embeddings $(\widetilde{\boldsymbol{h}}_u, \tilde{\boldsymbol{z}}_i)$ are concatenated into a single feature vector $x_{ui}$. This vector is then passed through multiple fully connected layers to capture higher-order interactions, with the final dense layer producing the predicted rating $\widehat{r_{ui}}$ on a 1 to 5 scale. The following subsections describe each of these components in detail.

### 3.1 Identifier Embeddings

User and item IDs are first encoded as one-hot vectors, $1_u \in \mathbb{R}^{|u|}$ for user $u$ and $1_i \in \mathbb{R}^{|i|}$ for item $i$, where |U| and |I| denote the total numbers of users and items in the dataset, respectively. These high-dimensional sparse vectors are then reduced to dense embeddings through linear projection matrices:

$$\mathbf{e}_u = W_U 1_u, \quad \mathbf{e}_i = W_I 1_i \tag{1}$$

where $W_U \in \mathbb{R}^{p \times |u|}$ and $W_I \in \mathbb{R}^{q \times |i|}$ are trainable parameters. The resulting dense vectors $\mathbf{e}_u \in \mathbb{R}^p$ and $\mathbf{e}_i \in \mathbb{R}^q$ represent the latent embeddings of the user and of the item, respectively. We map both user and item identifiers into the same latent space by setting $p = q$, ensuring that $\mathbf{e}_u$ and $\mathbf{e}_i$ are compatible with one another when concatenated in the prediction network.

### 3.2 Review Embeddings

Historical reviews are incorporated from two independent sources: the $k$ most recent reviews written by the user and the $k$ most recent reviews associated with the item. Each review is encoded into a fixed $d$-dimensional vector using SBERT, resulting in a sequence of embeddings for both user and item sides.

**User Reviews (Historical).** For each user $u$, let $\mathcal{R}_u^{(t)} = \{r_1, r_2, \ldots, r_k\}$ denote the $k$ most recent reviews authored by $u$ prior to timestamp $t$. Each review $r_j \in \mathcal{R}_u^{(t)}$ is encoded as $\boldsymbol{h}_{rj} \in \mathbb{R}^d$. The embeddings are concatenated into a single vector:

$$\overline{\mathbf{h_u}} = \left[ \mathbf{h}_{r_1} \parallel \mathbf{h}_{r_2} \parallel \ldots \parallel \mathbf{h}_{r_K} \right] \in \mathbb{R}^{k \cdot d} \tag{2}$$

This concatenated vector is then passed through a multi-layer perceptron (MLP), referred to as a Learning Layer in Figure 1, to reduce dimensionality, denoise, and highlight the most informative semantic features:

$$\widetilde{\boldsymbol{h_u}} = \phi\left( \text{MLP}_u(\overline{\boldsymbol{h_u}}) \right) \tag{3}$$

Where $\phi(\cdot)$ denotes the ReLU activation function.

**Item Reviews (Historical).** Analogously, for each item $i$, let $\mathcal{R}_i^{(t)} = \{r_1, r_2, \ldots, r_k\}$ be the $k$ most recent reviews associated with $i$ prior to $t$. Each review $r_j \in \mathcal{R}_i^{(t)}$ is encoded as $z_{r_j} \in R^d$. The embeddings are concatenated:

$$\overline{\boldsymbol{z_i}} = \left[ \boldsymbol{z}_{r_1} \parallel \boldsymbol{z}_{r_2} \parallel \ldots \parallel \boldsymbol{z}_{r_k} \right] \in R^{k \cdot d} \tag{4}$$

and then refined through a separate learning layer $\widetilde{\boldsymbol{z_i}} = \phi\left( \text{MLP}_i(\overline{\boldsymbol{z_i}}) \right)$.

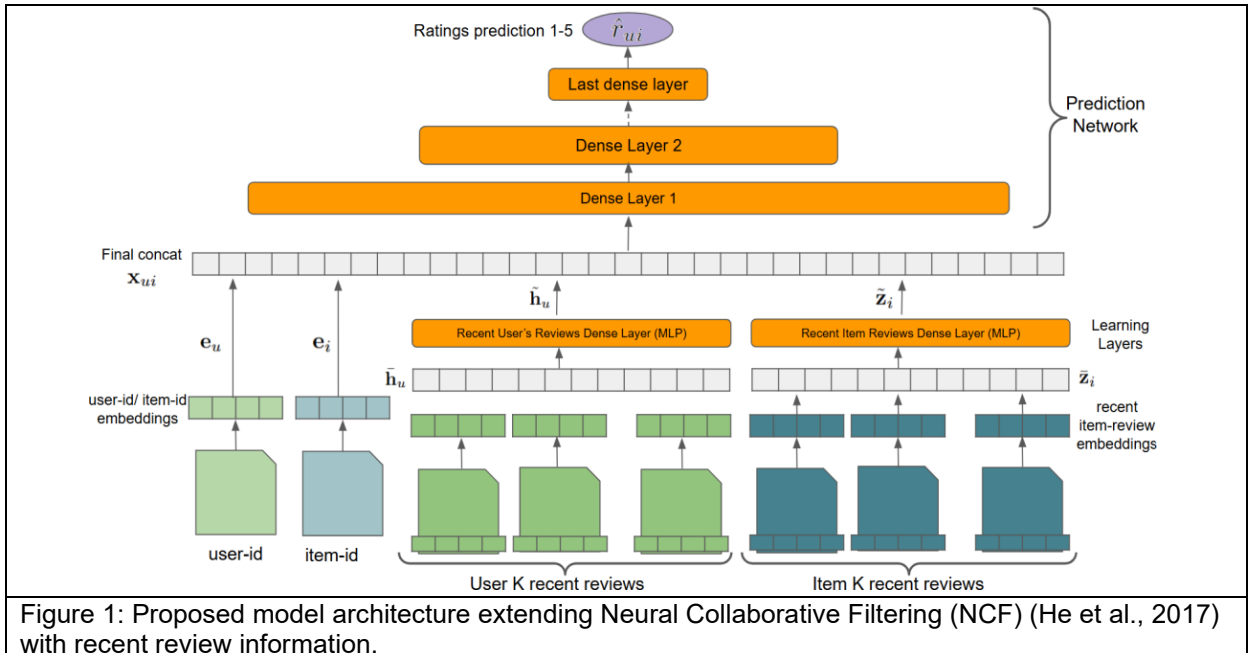

Figure 1: Proposed model architecture extending Neural Collaborative Filtering (NCF) (He et al., 2017) with recent review information.

Thus, the model produces two refined review representations: $\widetilde{\mathbf{h}_u}$ for user-side reviews and $\widetilde{z_i}$ for item-side reviews. These outputs of the learning layers are subsequently fused with the user and item ID embeddings in the prediction network.

### 3.3 Prediction Network

The final representation is formed by concatenating the identifier embeddings and the refined review embeddings:

$$x_{ui} = \left[\mathbf{e}_u \parallel \mathbf{e}_i \parallel \widetilde{\boldsymbol{h}_u} \parallel \widetilde{\boldsymbol{z}_i}\right] \tag{6}$$

This vector is passed through an MLP layer. The hidden layers are defined recursively as:

$$h^{(l)} = \sigma\left(W^{(l)} h^{(l-1)} + b^{(l)}\right), \quad l = 1, \dots, t \tag{7}$$

with $h^{(0)} = x_{ui}$. The final output layer produces the predicted rating:

$$\widehat{r_{ui}} = W^{(t+1)} h^{(t)} + b^{(t+1)} \tag{8}$$

The network is trained using mean squared error (MSE) loss, where $\mathcal{D}$ is the set of observed ratings:

$$\mathcal{L} = \frac{1}{|\mathcal{D}|} \sum_{(u,i) \in \mathcal{D}} (r_{ui} - \widehat{r_{ui}})^2 \tag{9}$$

To address our research questions, we apply the recommendation model under two main settings: one trained with historical human-authored reviews and another trained with historical reviews generated by an LLM (produced using different types of prompts, as further explained in Section 4). We additionally test cross train-test scenarios in which the proposed RS architecture was trained on human reviews but tested on AI-generated reviews, and vice versa. In all cases, the underlying architecture and the training procedure remain identical, allowing us to isolate the effect of review authorship on recommendation performance.

## 4   DATA

## 4.1   Human-Written Reviews

Our experiments are based on a dataset of hotel reviews from TripAdvisor, covering a ten-year period from September 2002 to December 2012. Each review includes both free-text content and a wide range of structured metadata (see Table 1 for field details). Because this dataset predates the introduction of LLM such as ChatGPT which was introduced to the general public in November 2022 (OpenAI, 2022), we can confidently state that all reviews were authored by humans, without any enhancement or generation by LLMs. The original dataset contained 79,180 reviews. After applying standard filters, keeping only English-language reviews from users and items with at least five interactions, the final dataset comprised 56,593 reviews, 6,920 unique users, and 1,969 unique hotels. Basic statistics for both the original human-authored reviews and the generated corpora are reported in Table 3. The distribution of hotel ratings is skewed toward positive sentiment: 40,848 reviews (72%) received 4–5 stars compared to only 15,745 reviews (28%) with 1–3 stars. This positivity bias is consistent with prior observations in online review platforms (Wen and Laporte, 2025; Shan et al., 2025) and may influence how recommender systems learn from the data (Zhou et al., 2025). Throughout this study, these human-authored reviews serve as the reference baseline in different scenarios evaluating the effects of AI-generated content.

| Table 1: Metadata fields available in the TripAdvisor dataset. | | |
|---|---|---|
| **Category** | **Field name** | **Description** |
| Review | Body text | Full text of the review |
| | Review date | Date the review was posted |
| | Stay date | Date of the hotel stay |
| | Overall rating | User's overall star rating (1–5) |
| | Multi-aspect rating (category rating) | Categories are varied and not mandatory and are not identical for each review. These include, but are not limited to: cleanliness, service, value, location, sleep quality, etc. Each category rating is also 1-5. |
| | Number of helpful votes | The total count of times other users marked a review as helpful. It reflects peer evaluation of usefulness, showing how many people found the |

|  |  | review valuable for their decision-making. |
|---|---|---|
| User | User ID | Unique identifier of the reviewer |
| Hotel | Hotel ID | Unique identifier of the hotel |
|  | Hotel name | Name of the hotel |
|  | Hotel location | Geographic location (city, country) |
|  | Category/class | Hotel star class or type |
|  | Hotel link | The website of the hotel |

## 4.2   AI-Generated Reviews from Existing Review Text or Metadata

As described in detail in the introduction, we assume that AI-generated content can be introduced through two main channels: (1) user-centric, where users employ LLM tools (e.g., ChatGPT, Gemini) to refine their own draft reviews; and (2) platform-centric, where the platforms generate and integrate synthetic reviews from available metadata into the RS pipeline.

To conduct a controlled assessment of how RS performance is affected by each class of AI-generated reviews, we generated, for each human-generated review in our dataset, a matching simulated user-centric review (based on the review text with some metadata) and a set of simulated platform-centric reviews (based only on metadata). We sought to generate these simulated reviews using natural and realistic prompts. Accordingly, when generating text with the LLM, we provided two types of instructions: a system-role message and a user-role message. The system-role message defines the model's overarching role and behavior, ensuring consistency in tone and task execution. The user-role message, in contrast, provides the specific input for each case, i.e., the text of a review we want to refine (for the user-centric scenario), and/or the review/hotel metadata (for the platform-centric scenario).

Below, we first describe how we generated textual reviews for the user-centric scenario (RQ2), followed by the platform-centric scenario (RQ3). All textual reviews were generated using OpenAI's GPT-4o-mini model, accessed via API with default parameters. We applied the same configuration to generate each corpus of reviews to ensure consistency in generation and reduce stylistic variance.

### 4.2.1 Generating User-Centric Reviews (RQ2)

In the user-centric scenario, consumers use LLMs to refine or enhance their own textual reviews. To simulate this scenario, we mimic a usage type in which users employ an AI assistant (such as ChatGPT, Claude, or Gemini) to refine their own original human-written text. Accordingly, the prompting strategy for creating the GenAI reviews was designed to improve existing user input rather than generate reviews solely from metadata. Specifically, we provided the GPT model with the original human review text, together with the overall rating assigned by the user and the category-specific ratings, along with instructions to enhance the text.

The system-role message used was:

> *You are a helpful assistant whose sole task is to assist the user in writing an overall textual review for a hotel they stayed at. Length should be average review length for hotels.*

And the user-role message was:

> *I am writing a review for TripAdvisor about my stay at [hotel-name]. I gave the hotel an overall rating of [overall-ratings] out of 5. Additionally, I gave the following ratings to specific criteria: [details-ratings]. Below is the draft of my review. Please improve it so that it is coherent, suitable for posting as a hotel review, and aligned with the star rating I gave. Draft: [user-review-text]*

### 4.2.2 Generating Platform-Centric Reviews (RQ3)

In the platform-centric scenarios, the platform itself generates review content directly from structured metadata, without any human draft. We simulated these scenarios as follows. The LLM was given the following system-role message:

> *You are a helpful assistant whose sole task is to assist the user in writing an overall textual review for a hotel they stayed at*
> *You will receive the following parameters in a JSON structure:*
> *1. **Score** - A JSON object containing the experience rating given by the user for each review category using a star rating system, typically from 1 (poor) to 5 (excellent).*

*The structure of the JSON will generally be:*
*{'service': [score], 'cleanliness': [score], 'overall': [score], 'value': [score], 'location': [score], 'sleep_quality': [score], 'rooms': [score] }.*
*Note - There might be additional/less categories.*
*2. **Location** - A JSON object containing the location of the hotel. The structure of the JSON will be:*
*{'region': [value], 'street-address': [value], 'postal-code': [value], 'locality': [value]}.*
*Note: The structure of this JSON object may vary slightly.*
*3. **Name** - The name of the hotel.*
*4. **Link** - The URL to the hotel's page on TripAdvisor.*
*5. **Date_stayed_in_hotel** - The date when the user stayed at the hotel, provided in the format <month year>.*
*6. **Date_review** - The date the review is being written.*
*7. **Class** - The hotel's class (from 1 to 5, with 5 being the highest).*
*Using these parameters, you will generate a review and return the review text.*
*Note - Do not specifically mention the star ratings in the review.*
*Refrain from explicitly entering the hotel website.*
*Length should be average review length for hotels.*

The LLM was then given a user-role message, which provides the case-specific input in the form of a JSON object containing the review ratings and hotel metadata on which the model generates the review. For example:

*{ "Score": {"service": 5, "cleanliness": 4, "overall": 5},*
*  "Location": {"region": "California", "locality": "San Diego"},*
*  "Name": "Pacific View Hotel",*
*  "Link": "https://tripadvisor.com/...",*
*  "Date_stayed_in_hotel": "June 2012",*
*  "Date_review": "July 2012",*
*  "Class": 4 }*

This setup reflects a prompt that draws on all the structured information available to the platform at the time of review generation. The model was instructed to compose a complete review by leveraging the full set of metadata: overall score, category-specific ratings, hotel name, location, link, stay dates, review date, and hotel class. The emphasis here is on producing a comprehensive review grounded in available signals, without additional stylistic guidance beyond faithfully rendering the underlying information.

In addition to this setup for generating platform-centric GenAI reviews, we created three additional datasets by modifying the system message to impose different stylistic tones. We focused on three types of tones: (i) encouraging, (ii) constructive, and (iii) critical, because they represent distinct framing strategies: strongly positive framing (encouraging), balanced and fair feedback (constructive), and explicitly negative framing (critical). Together, they capture a meaningful spectrum of how GenAI reviews might influence user preferences and RS learning.

These variants followed the same metadata structure but added a short instruction to the system message to shape the generated text according to the designated tone. Table 2 summarizes the differences in the system message instructions across the four platform-centric tone variants.

Table 2: System message instructions provided to the LLM for generating reviews in four platform-centric tone variants.

| Variant | Instruction |
|---|---|
| Neutral | Using these parameters, you will generate a review and return the review text. |
| Encouraging | Using these parameters, generate a warm and enthusiastic review that highlights the most positive aspects of the experience. Use a friendly and appreciative tone. |
| Constructive | Using these parameters, generate a fair and thoughtful review that reflects both the strengths of the experience and areas that could be improved. Use a balanced and constructive tone. |
| Critical | Using these parameters, generate a candid and critical review that emphasizes areas where the experience did not meet expectations. When appropriate, use a direct and dissatisfied tone while remaining factual and respectful. |

Together, this procedure resulted in four platform-centric datasets: one neutral platform-centric dataset (built with the neutral prompt and used in all analyses of RQ3), and three stylistically modified datasets (used in RQ3.c). These corpora allow us to test whether changing the tone of generated reviews influences RS performance.

Table 3 summarizes the basic statistics for each dataset. The table reports the varying text-level characteristics, including average word count, average character count, and total vocabulary size (total unique word count across all reviews).

Table 3: Text-level dataset statistics: average word count, average character count, and total vocabulary size across conditions.

| Data Set | Avg. Word Count | Avg. Char Count | Total Vocab Size |
|---|---|---|---|
| Original Human Text | 187.98 | 1023.19 | 193,493 |
| User-Centric | 265.26 | 1593.26 | 103,269 |
| Platform-Centric: Neutral | 200.98 | 1230.58 | 18,299 |
| Platform-Centric: Constructive | 258.74 | 1601.36 | 20,909 |
| Platform-Centric: Encouraging | 216.09 | 1290.65 | 21,378 |
| Platform-Centric: Critical | 264.52 | 1636.79 | 28,741 |

Notably, as shown in Table 3, human and user-centric reviews show much larger vocabularies than platform-centric variants, while constructive and critical reviews are considerably longer. These characteristics are consistent with previous literature showing that human texts tend to have richer vocabularies, as discussed in Section 2.1.

## 5 METHODOLOGY

### 5.1 RQ1: Systematic Differences between Human- and AI-Generated (User- or Platform-Centric) Reviews

Our analysis for RQ1 systematically characterizes textual differences between human-generated reviews and LLM-generated reviews in both user-centric and platform-centric scenarios. For the latter, we focus on the "neutral" platform-centric corpus (Table 3). Our analyses focus on several key features. First, we assess internal similarity, defined as the extent to which reviews within a particular corpus (AI-generated vs. human-written) are similar to one another (this evaluation is denoted RQ1.a). Second, we evaluate textual features that capture the inherent properties of the content itself, focusing on lexical diversity, emotional expression, and sentiment (denoted RQ1.b–d, respectively). Importantly, while internal similarity is computed using embedding representations of the reviews, all other textual features are calculated directly from the review text. Below we describe the operationalization of each of the four features (see Table 4 for a summary). All analyses were performed on a random sample of 1,000 observations, and, for robustness, subsequently repeated on larger samples of 5,000 and 20,000 observations. We conduct each comparison separately for each AI-generated corpus (the user-centric corpus and the neutral platform-centric corpus) against the original human-written reviews.

| Table 4: Feature operationalization for RQ1. | | | |
|---|---|---|---|
| Sub-RQ | Feature | Method | Description |
| RQ1.a | Internal Similarity | SBERT Embedding[1]-Level | Cosine similarity between SBERT embeddings of the textual reviews |
| RQ1.b–d | Lexical Diversity | Word-Level Text | Unique words in the original review text, normalized by review length (total words). |
| | Emotion Profile | Pretrained Emotion Model | Distribution of dominant emotions in reviews, calculated on the original review text |
| | Sentiment Polarity | VADER Sentiment | Mean scores across sentiment classes, calculated on the original review text |

To assess internal similarity, which reflects the degree of linguistic homogeneity within a corpus and thus the informational diversity available for algorithmic learning, we represent each review as a high-dimensional vector using SBERT embeddings. We then calculate pairwise cosine similarities between all unique pairs of reviews within each corpus (human-generated, user-centric AI-generated, or platform-centric AI-generated), where values range from –1 (opposite direction) through 0 (no relation) to 1 (identical direction, complete similarity). We then conduct two separate statistical tests: one comparing the internal similarity of the user-centric corpus against human-written reviews, and another comparing the platform-centric corpus against human-written reviews. Lower internal similarity between pairs indicates more diverse texts, whereas higher similarity reflects more standardized patterns.

For lexical diversity, we compute the ratio of unique words to the total number of words (excluding stopwords) to capture richness and variability in the vocabulary and expression of the texts. Formally, lexical diversity for review $i$ is defined as: $\frac{\text{UniqueWords}_i}{\text{TotalWords}_i}$. As with internal similarity, we perform a separate comparison between each AI-generated corpus and the human-

---

written corpus. Since each human review has a corresponding AI-generated review in each corpus, we use paired t-tests for these comparisons.

For emotional expression, we first apply the RoBERTa-based go emotions model[2] to identify the dominant emotion in each (human- or LLM-generated) review. Specifically, this is a pre-trained multi-label emotion classification model, which predicts probabilities across 28 emotion categories; for each review, we take only the dominant predicted emotion, which will characterize this review. We then compare the distribution of these emotion categories between human-written and AI-generated reviews, conducting the analysis separately for the user-centric and platform-centric corpora. This approach enables us to capture potential differences in both the range and concentration of emotional content across groups.

For sentiment, we use the Valence Aware Dictionary and Sentiment Reasoner (VADER; Hutto et al. 2014[3]) to assign three polarity scores to each review: positive, neutral, and negative. VADER is a recognized tool for sentiment analysis in IS literature (Liu et al, 2020, Przegalinska et al 2025). We examine each of these polarity dimensions separately. We use paired t-tests for each polarity score to compare human-written reviews with their user-centric and platform-centric AI-generated counterparts. This enables us to identify systematic differences in sentiment distribution between human-generated reviews and the AI-generated reviews in each corpus.

Taken together, these four measures will provide a comprehensive comparison of human- and (different classes of) AI-generated reviews.

## 5.2   RQ2: Effects of User-Centric GenAI Reviews on RS Performance

---

In the user-centric scenario, users employ GenAI tools to refine their own reviews, raising questions about how such AI-generated content affects RSes that were originally designed to rely on human-authored opinions. Our analysis for RQ2 entails the following settings and procedures, which we use to compare RS performance on human-generated reviews to performance on user-centric AI-generated reviews (see Section 4.2.1).

**Recommendation Evaluation Metrics**. We use two complementary sets of metrics to assess RS recommendation performance. First, we follow the evaluation protocol of He et al. (2017) and use standard metrics to evaluate the accuracy of predicted ratings and to assess ranking quality. For rating prediction, we evaluate Root Mean Square Error (RMSE) and Mean Absolute Error (MAE). RMSE represents the square root of the average squared difference between predicted and true ratings, while MAE measures the average absolute deviation between predicted and true ratings across all user–item pairs in the test set. For rankings, we use two common metrics to assess the quality of top-k recommendations: (1) Mean Reciprocal Rank (MRR@k), the average reciprocal rank of the first relevant item in the top-k list; and (2) Normalized Discounted Cumulative Gain (NDCG@k), which measures ranking quality by rewarding hits at higher positions. All ranking metrics are computed at multiple list lengths of k (3, 5, 10, and 20). We report results with k = 10 (results for the other k's were consistent) and use the same protocol for generating and evaluating recommendation lists as in He et al. (2017).

Second, in addition to the standard recommendation metrics, we examine business-oriented outcomes, including item helpfulness and item quality (i.e., hotel star ratings), to understand how AI-generated reviews influence the business relevance of recommendations. Specifically, we evaluate: (1) Average Hotel Stars@k, the average hotel class (1–5 stars) among the top-k recommendations, used as a proxy for quality-oriented ranking. (2) Average Popularity@k, the average popularity rank of items among the top-k recommendations, measured by the number of reviews each hotel has received. This metric captures the extent to which the recommender favors already-popular items over less-reviewed ones. (3) Average Helpfulness@k, the average number of useful votes for reviews linked to top-k recommendations. This metric reflects how useful those reviews are perceived to be by other users. (4) Average Regional Spread@k, the average number of distinct regions represented among the top-k recommended items. In this context, a region corresponds to a hotel's state or province. This metric captures how geographically dispersed the recommendations are: lower values indicate concentration in fewer regions, whereas higher values reflect broader geographic diversity across regions.

**Data Splitting Strategies.** To generate recommendation predictions for evaluation, we adopt the same leave-one-out strategy as He et al. (2017). Specifically, for each user, we hold out their latest interaction as the test set and use the remaining interactions for training. This setup reflects real-world recommendation scenarios, where the system predicts future user preferences based on their historical behavior rather than global chronological order. For ranking evaluation, we follow the common strategy in He et al. (2017), randomly sampling 99 items not interacted with by the user and ranking the test item among them. Additionally, for the ranking procedure, we restrict the test set to positive reviews only (i.e., five-star ratings). To ensure consistency across all experimental conditions, we use identical train/test splits across all datasets.

**Model Comparison.** We evaluate three RS models using the proposed architecture in Section 3. The first baseline is the NCF model (NCF) (He et al., 2017), which includes only user and item identifiers without textual reviews. The second model is NCF-Human, which incorporates both user and item identifiers and historical human-authored reviews (see Section 4.1). The third model is NCF-User-Centric, which incorporates user-centric GenAI reviews (see Section 4.2.1).

**Training Setup and Hyperparameter Tuning.** We used a validation set to tune hyperparameters for each RS baseline model's training, with final performance comparisons conducted on the test set. Because data splitting is done per user, we selected 10% of each user's interactions that occurred after the training window (but before the interaction time in the test set) as validation data. This setup simulates a realistic temporal recommendation pipeline while preserving user-level consistency.

We tested multiple latent sizes of user and item representations [20,50, and 100] and report the best results obtained with the size of 100. Model parameters were initialized from a Gaussian distribution with mean 0 and standard deviation 0.01. Each RS model was optimized with MSE loss using the Adam optimizer. Batch sizes of [128, 256, 512, 1024] and learning rates of [0.0001, 0.0005, 0.001, 0.005] were tested, and we report results with learning rate of 0.0005. For review representations, we used the traditional SBERT model to embed review content. This model produces 384-dimensional vectors optimized for sentence-level semantic similarity, offering a good balance between efficiency and accuracy. For the network prediction layers, we evaluated the reduction proportions of [0.75, 0.5, 0.3, 0.25] to control the number of neurons in successive layers. A reduction proportion specifies how the layer size decreases relative to the previous one (e.g., a proportion of 0.5 means that each subsequent layer has half the number of neurons of its predecessor). We report results using the best-performing value of 0.25. All models were trained for a fixed 50 epochs to ensure consistency in training duration.

This hyperparameter tuning procedure was applied consistently to each RS model (NCF, NCF-Human, and NCF-User-Centric) to ensure a fair comparison under different conditions.

The recommendation framework was implemented in Keras with a TensorFlow backend. All models were trained under tightly controlled and reproducible settings, including fixed random seeds for NumPy, Python, and TensorFlow.

**Significance Testing**. We apply paired t-tests on model prediction errors to determine whether performance differences between models trained on human-authored versus AI-generated reviews are statistically significant. Tests are conducted on matched sets to ensure that comparisons isolate the effect of review source rather than sample variation. For example, in the case of RMSE, we apply the t-test on the squared prediction errors of each entry in the test set under both training conditions we want to compare, and we use a similar procedure for the other evaluation metrics (e.g., MAE, NDCG@k, MRR@k).

### 5.3 RQ3: Effects of Platform-Centric Reviews on RS Performance

As discussed previously, RQ3 evaluates RS performance in scenarios where reviews are not authored by users but are generated entirely from structured metadata (e.g., multi-criteria ratings and hotel attributes) (see Section 4.2.2). RQ3 consists of multiple parts (RQ3.a,b,c). Unless otherwise noted, all experimental settings in RQ3 are identical to those in RQ2. In what follows, we highlight the methodological aspects specific to each part.

**RQ3.a: Platform-centric baseline comparison**. In the first part of RQ3, we introduce an NCF-Platform-Centric model trained on reviews generated directly from structured metadata using the neutral prompt (Section 4.2.2). We compare this model against the two established RQ2 baselines: NCF (no reviews) and NCF-Human (human-authored reviews) (Section 5.2). All hyperparameter selection follows the same procedure as in RQ2 (Section 5.2), holding every aspect of the pipeline constant except the source of review text.

**RQ3.b: Implications of platform-centric reviews under cross train-test conditions**. In the second part of RQ3, we evaluate whether RSes trained on one type of review can generalize when tested on another. To do this, we construct two cross train-test conditions. In the Human→GenAI setting, models are trained on human-authored reviews but evaluated on test data where user histories are represented by the corresponding platform-centric AI-generated reviews. In the GenAI→Human setting, models are trained on platform-centric reviews but evaluated on human-authored ones. To ensure comparability, all other experimental elements remain unchanged from RQ2 (Section 5.2), including the data splits, evaluation metrics, and embedding model. We also fix the model configuration to the final hyperparameter values selected in RQ2, rather than conducting a new hyperparameter search. This approach ensures that observed differences reflect cross train-test effects rather than changes in training procedure.

**RQ3.c: Implications of tone-based framing strategies for RSes**. In the third part of RQ3, we investigate whether the tone of platform-generated reviews influences RS outcomes. To do this, we build three additional NCF-Platform-Centric models trained on reviews generated with distinct stylistic instructions: Constructive, Encouraging, and Critical. Together with the Neutral platform-centric model (baseline), this yields four platform-centric variants (See Table 2).

## 6   RESULTS

### 6.1   RQ1: Systematic Differences between Human- and AI-Generated (User- or Platform-Centric) Reviews

We now turn to the results of RQ1, comparing human-written reviews with both user-centric and platform-centric AI-generated reviews across four textual dimensions. Results for the four textual dimensions are presented, respectively, in Figures 2, 3, 4, 5 side by side, with the left panel showing Human vs. User-Centric AI reviews and the right panel showing Human vs. Platform-Centric AI reviews. Due to run-time limitations of the emotions model, results are presented for a random sample of 1000 observations. For robustness, we reran the analyses on larger samples of 5,000 and 20,000 observations for all measures other than emotions. The results were consistent.

**Internal similarity**: Cosine similarity distributions, computed from SBERT embeddings, are displayed in Figure 2. Human-authored reviews exhibit greater variability, with lower average similarity across review pairs (mean = 0.47). By contrast, both user- and platform-centric AI reviews show higher internal similarity, indicating more standardized and homogeneous generation outcomes (means = 0.51 and 0.57, respectively). Statistical tests (t-tests) confirm that these differences between the groups are significant in both scenarios (user-centric vs. human: t = −199.8, p < 0.001; platform-centric vs. human: t = −556.1, p < 0.001).

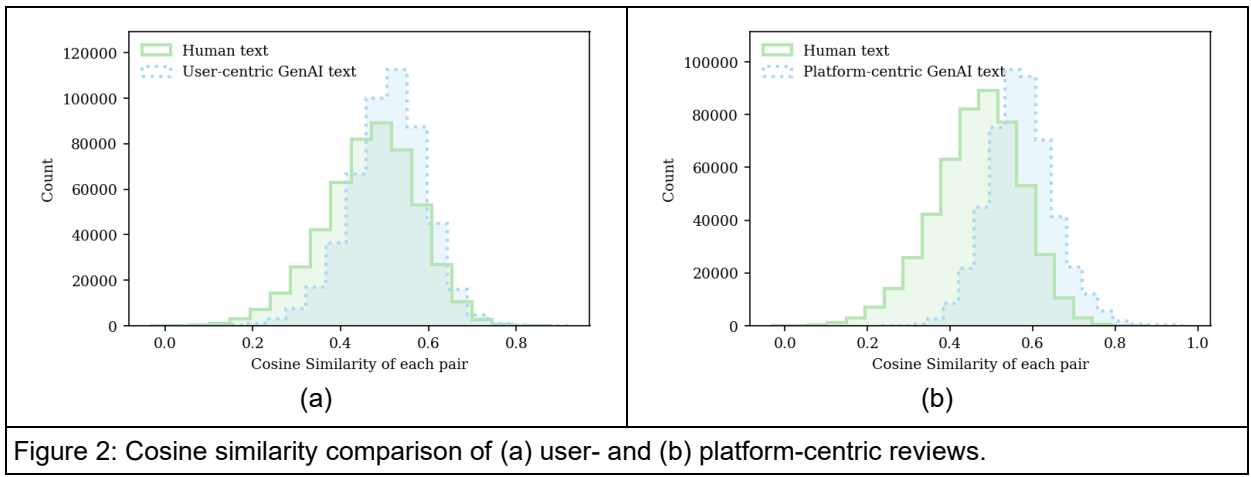

Figure 2: Cosine similarity comparison of (a) user- and (b) platform-centric reviews.

**Lexical diversity**: Figure 3 shows the distributions of lexical diversity, measured as the ratio of unique words to total words. Human-written reviews consistently achieve higher lexical diversity than AI-generated reviews, reflecting a richer and more varied vocabulary. On average, human reviews have a mean lexical diversity of 0.872 (SD = 0.075). By contrast, user-centric AI reviews exhibit a lower mean of 0.864 (SD = 0.035), while platform-centric AI reviews show an even lower mean of 0.842 (SD = 0.033). Beyond these differences in means, human-authored reviews also display greater variability compared to AI-generated reviews, whose lexical diversity values are more tightly clustered. Paired t-tests confirm that the mean differences are statistically significant in both scenarios (user-centric vs. human: t = 2.81, p < 0.005; platform-centric vs. human: t = 11.59, p < 0.001).

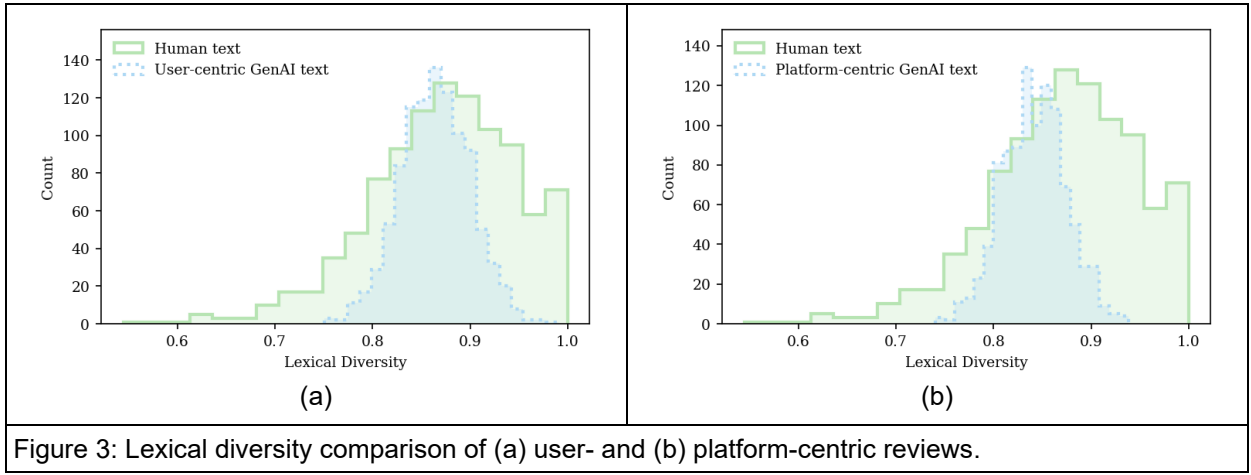

Figure 3: Lexical diversity comparison of (a) user- and (b) platform-centric reviews.

**Emotional expression**: The distribution of dominant emotions, predicted by the go emotions model, is shown in Figure 4. Human reviews span a wide emotional range, with 22 of the 28 emotion categories represented. In contrast, AI-generated reviews cluster around a narrower set of categories, especially admiration, which dominates the distribution. User-centric AI reviews include 14 emotion categories, while platform-centric AI reviews are even more limited, covering only 9. This pattern points to a narrower emotional range in AI-generated reviews compared to the broader variety found in authentic human-authored ones.

**Sentiment:** Sentiment polarity score means, derived from the VADER model, are presented in Figure 5. AI-generated reviews tend to be slightly more positive, less neutral, and less negative than their human counterparts. To assess differences, we ran paired t-tests separately for each polarity dimension (positive, neutral, negative) and for each scenario (user-centric vs. human; platform-centric vs. human). The results confirm significant differences between human and AI reviews: positive (user-centric: t = −15.18, p < 0.001; platform-centric: t = −21.78, p < 0.001),

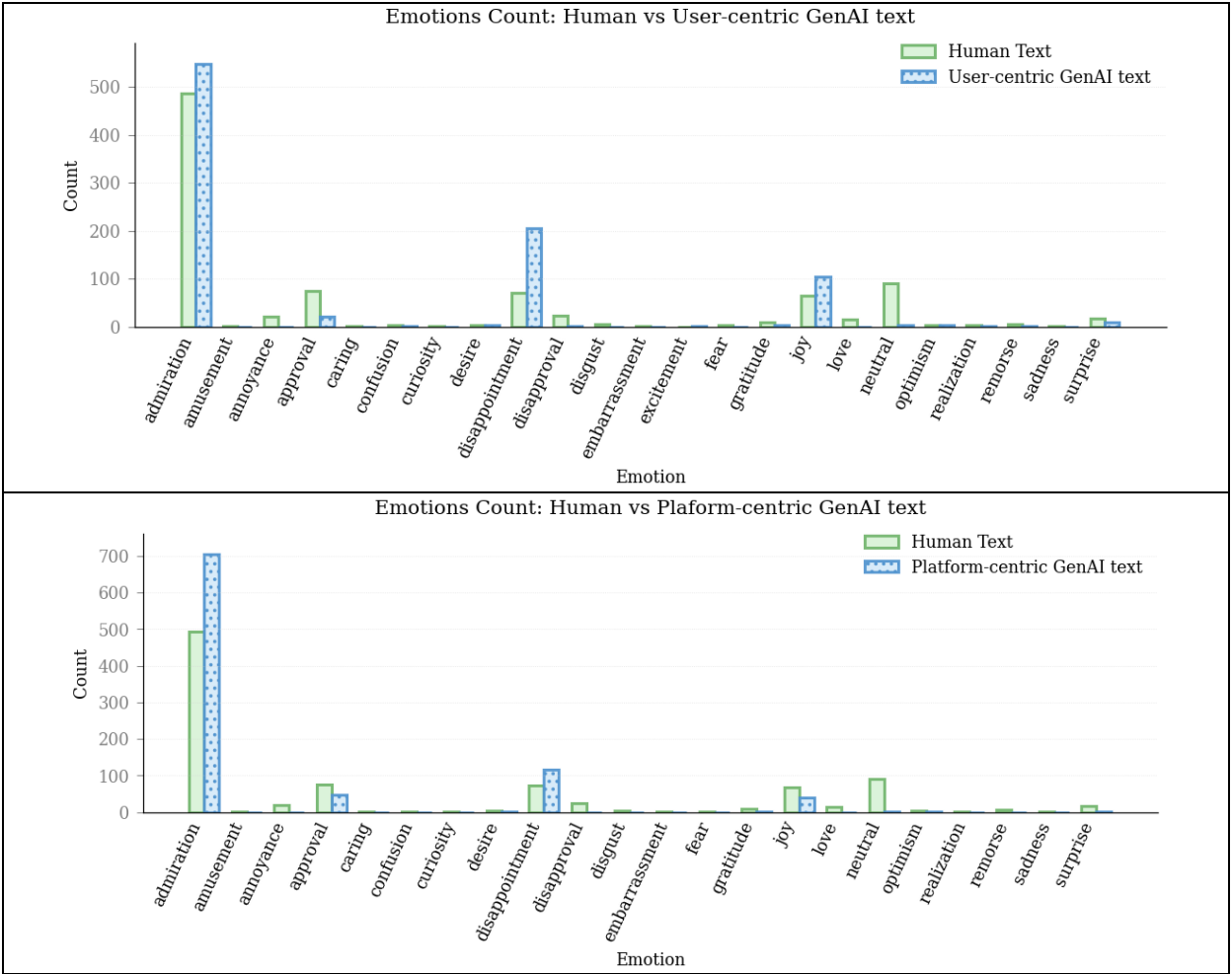

Figure 4: Emotion distribution comparison of user- and platform-centric reviews.

neutral (user-centric: t = 9.30, p < 0.001; platform-centric: t = 10.59, p < 0.001), and negative

(user-centric: t = 12.07, p < 0.001; platform-centric: t = 27.42, p < 0.001).

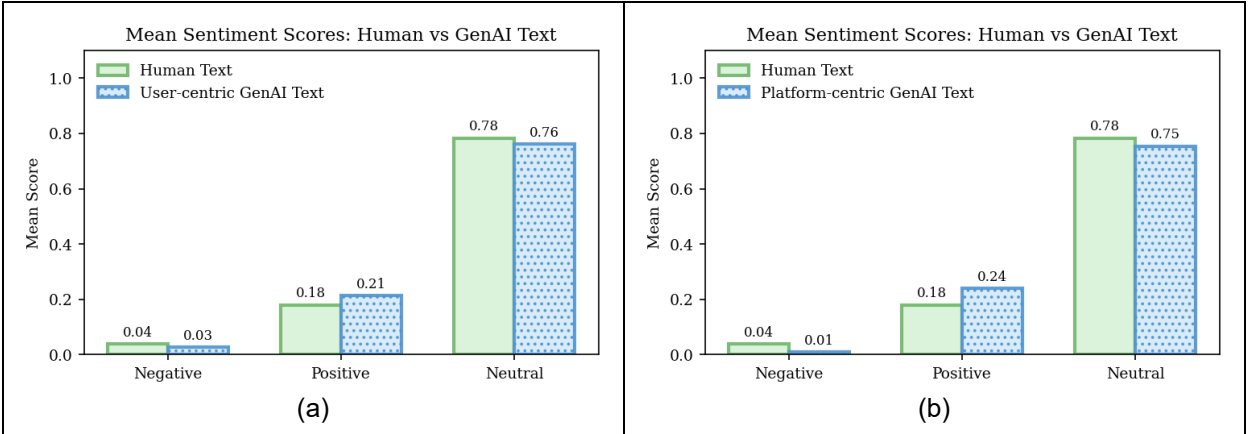

Figure 5: Sentiment polarity comparison of (a) user- and (b) platform-centric reviews.

Taken together, these results demonstrate consistent and systematic differences between human-generated and AI-generated reviews, strengthening the finding of previous literature. Human-authored reviews display greater lexical richness, emotional variety, and variability in expression, whereas AI-generated reviews are more uniform, with narrower emotional ranges, tending to be slightly more positive and less neutral or negative. These differences have important implications for how recommender systems interpret and integrate user feedback, particularly in environments with both human and AI-generated content. As such, they make the following evaluation of the integration of GenAI reviews into recommendation systems both necessary and valid.

### 6.2   RQ2: Effects of User-Centric GenAI Reviews on RS Performance

To address RQ2, we trained a traditional deep-learning NCF recommendation model containing reviews, as described in Section 3, and compared original human-written reviews with their user-centric AI-generated counterparts to evaluate their impact on recommendation performance.

Table 5 presents the results. The asterisks in the table denote the statistical significance of paired t-tests comparing the NCF-Human model (i.e., NCF trained on original human-written reviews) with the NCF-User-Centric model (i.e., NCF trained on AI-generated reviews created using our user-centric prompts). Across all metrics, NCF-Human performs significantly better than NCF-User-Centric, underscoring the importance of the data on which the model is trained: models trained on original human reviews capture user preferences more effectively and deliver higher-quality recommendations.

Importantly, while NCF-Human performs better than NCF-User-Centric, both substantially outperform the baseline NCF model with no reviews. This finding is consistent with prior work (Fu et al., 2019), which shows that integration of human textual data improves performance by enriching sparse interaction histories with semantic content. At the same time, our results extend this literature by demonstrating that not only human-authored but also AI-refined textual data can yield meaningful improvements, highlighting how generative text contributes to recommender performance. For NCF-Human, the improvements are statistically significant across all metrics: RMSE decreases from 1.154 to 1.014 (a 12.1% reduction), MAE from 0.888 to 0.781, MRR@10 increases from 0.061 to 0.078 (a 27.9% improvement), and NDCG@10 from 0.094 to 0.114 (a 21.3% improvement). For NCF-User-Centric, the gains are statistically significant for the rating metrics, with RMSE improving from 1.154 to 1.029 (a 10.8% reduction) and MAE from 0.888 to 0.798, while ranking metrics also improve but do not reach statistical significance. These improvements, whether the larger 12.1% reduction achieved by the NCF-human model or the more modest 10.8% reduction achieved by the NCF-User-Centric model, both exceed the 1–2% gains that have historically translated into measurable lifts in downstream KPIs in large-scale A/B tests (e.g., Netflix Prize, Amazon) (Smith and Linden, 2017). These results suggest that even the smaller AI-driven improvements are practically meaningful when scaled to millions of recommendations.

| Model | Rating Metrics | | Ranking Metrics | |
|---|---|---|---|---|
| | RMSE | MAE | MRR@10 | NDCG@10 |
| NCF | 1.154 | 0.888 | 0.061 | 0.094 |
| NCF-Human | **1.014 \*\*\*** | **0.781\*\*\*** | **0.078\*\*** | **0.114\*\*** |
| NCF-User-Centric | 1.029 | 0.798 | 0.067 | 0.102 |

Table 5: Performance metrics comparison between the baseline NCF model, original human reviews, and user-centric (AI-corrected) reviews.

Note: * p ≤ 0.1, ** p ≤ 0.05, *** p ≤ 0.001 for the comparison between NCF-Human and NCF-User-centric

Similarly to the rating- and ranking-based performance metrics above, the business-oriented outcomes also show a consistent pattern: both the NCF-Human and the NCF-User-Centric models outperform the baseline NCF, but the NCF-Human model holds an overall advantage. Table 6 presents the results, with asterisks denoting the statistical significance of paired t-tests comparing the NCF-User-Centric to NCF-Human. Across the measures, NCF-Human produces significantly higher average hotel stars (3.813 vs. 3.789) and higher popularity exposure (27.632 vs. 26.117). It also yields a narrower regional spread (7.497 vs. 7.602), while helpfulness, measured as the average number of "useful" votes for reviews linked to top-10 recommendations, is slightly higher (1.405 vs. 1.401) but not significant. These findings suggest that recommendation models trained on human-authored reviews surface higher-quality hotels and more popular items, and that they tend to produce geographically more focused recommendation patterns. In contrast, user-centric GenAI reviews disperse recommendations across a broader set of regions, a difference that may reflect a tradeoff between focus and diversity rather than a clear performance advantage, depending on the platform's goals. On helpfulness, we see that the user-centric GenAI model comes very close to human performance, indicating that models trained on GenAI reviews surface recommendations that are similar in informativeness to those trained on human-authored content.

Importantly, overall, both the NCF-Human and NCF-User-Centric models improve business outcomes compared to the no-review baseline. For NCF-Human, the improvements are significant across hotel quality (3.813 vs. 3.740), popularity (27.632 vs. 25.151), and helpfulness (1.405 vs. 1.379), whereas the reduction in regional spread (from 7.542 to 7.497) is marginally significant ($p < 0.1$). These results confirm that incorporating human-authored reviews (vs. no reviews) enables the recommender to surface recognized, higher-quality hotels associated with reviews that users find more useful. In turn, the NCF-User-Centric model also achieves significant improvements over the NCF baseline across the measures of hotel quality (3.789), popularity (26.117), and helpfulness (1.401); moreover, it produces a slightly wider regional spread of 7.602. These findings show that user-centric GenAI reviews also help the system highlight higher-quality hotels and expand exposure (vs. no reviews), though they shape coverage differently by distributing recommendations across a broader geographic range.

Table 6: Business metrics comparison between original human reviews, user-centric GenAI reviews and the NCF baseline.

| Model | Avg Popularity@10 | Avg Helpfulness@10 | Avg Regional Spread@10 | Avg Hotel Stars@10 |
|---|---|---|---|---|
| NCF | 25.151 | 1.379 | 7.542 | 3.740 |
| NCF-Human | **27.632***** | **1.405** | **7.497**** | **3.813***** |
| NCF-User-Centric | 26.117 | 1.401 | 7.602 | 3.789 |

Note: *$p \leq 0.1$, ** $p \leq 0.05$, *** $p \leq 0.001$ for the comparison between NCF-Human and NCF-User-centric

Taken together, the results across both performance and business-oriented outcomes present a consistent and coherent pattern. Human-authored reviews provide the strongest improvements overall: they reduce rating error, improve ranking quality, and at the same time enhance business outcomes by promoting higher-quality items, surfacing more helpful reviews, and focusing recommendations on relevant geographic areas. User-centric GenAI reviews, while not as strong, nonetheless deliver significant improvements compared to the NCF baseline. They come close to human-authored reviews on helpfulness and offer distinct advantages such as broader geographic coverage, which may be desirable in certain contexts. These findings underscore that human-authored reviews remain the gold standard for personalization and strategic platform goals, but user-centric GenAI reviews provide a meaningful and practical substitute in settings where human content is limited.

### 6.3   RQ3: Effects of Platform-Centric Reviews on RS Performance

RQ2 examined user-centric models trained on human-written reviews refined by GenAI. RQ3 shifts the focus to platform-centric scenarios where the platform generates the entire review corpus using GenAI (see Section 4.2.2 for details on this corpus construction). Recall that in this setting, platforms create reviews directly from interaction data, such as hotel information and user ratings, rather than relying on content written by users.

We examine three aspects of platform-centric review generation. First, we apply RQ2's methodology using a fully AI-generated review corpus to establish the effects of these reviews on performance and business outcomes (RQ3.a). Second, we test cross-training scenarios where models are trained on one review source (human or AI) but tested on another, examining how training-test misalignment affects outcomes (RQ3.b). Third, we evaluate whether tone-based framing strategies, using encouraging, constructive, or critical tones, can enhance the effectiveness of AI-generated reviews when integrated into RSes (RQ3.c). Together, these analyses assess how platform design choices in generating and deploying AI reviews influence both RS performance and business outcomes.

### 6.3.1 RQ3.a: Platform-Centric Reviews: Implications for Performance and Business Outcomes

For RQ3.a, similarly to RQ2, we compared the performance of our NCF-Human model to that of the NCF-Platform-Centric model (based on platform-centric GenAI reviews produced using the neutral prompt), and we evaluated both against the NCF baseline with no reviews. It is important to note that the NCF and NCF-Human models here are identical to those presented in RQ2, so the comparisons between NCF-Human and NCF performance replicate the same significant results described in the previous section. Table 7 presents the performance results, with asterisks denoting the statistical significance of paired t-tests comparing NCF-Platform-Centric to NCF-Human. Across all metrics, NCF-Human performed significantly better, again underscoring the richness of authentic user text for capturing fine-grained preferences and delivering superior recommendation accuracy. For example, RMSE drops to 1.014 for NCF-Human compared to 1.052 for NCF-Platform-Centric, and ranking metrics show sizable gaps as well (MRR@10 = 0.078 vs. 0.063, NDCG@10 = 0.114 vs. 0.099).

When compared to the baseline NCF model, NCF-Platform-Centric demonstrates meaningful improvements, albeit more modest than those achieved by the NCF-Human model (elaborated in section 6.2). Specifically, compared with NCF, NCF-Platform-Centric achieves significantly better rating metrics, with RMSE decreasing from 1.154 to 1.052 (an 8.8% reduction) and MAE from 0.888 to 0.821. Ranking metrics improve directionally (MRR@10 = 0.063 vs. 0.061, NDCG@10= 0.099 vs. 0.094), but the differences do not reach statistical significance. This suggests that platform-generated reviews inject useful semantic signals that improve predictive accuracy, but not enough to consistently enhance ranking precision.

| Table 7: Performance metrics comparison between original human reviews, the platform-centric reviews with neutral tone and the NCF baseline. | | | | |
|---|---|---|---|---|
| **Model** | **Rating Metrics** | | **Ranking Metrics** | |
| | **RMSE** | **MAE** | **MRR@10** | **NDCG@10** |
| NCF | 1.154 | 0.888 | 0.061 | 0.094 |
| NCF-Human | **1.014 \*\*\*** | **0.781\*\*\*** | **0.078\*\*** | **0.114\*\*** |
| NCF-Platform-Centric | 1.052 | 0.821 | 0.063 | 0.099 |

Note: * p ≤ 0.1, ** p ≤ 0.05, *** p ≤ 0.001 for the comparison between the NCF-Human and NCF-Platform-Centric

Table 8 presents the business-oriented outcomes. Once again, NCF-Human mostly outperforms both NCF and NCF-Platform-Centric, with significantly higher popularity (27.632 vs. 26.425), helpfulness (1.405 vs. 1.377), and hotel stars (3.813 vs. 3.764), as well as a narrower regional spread (7.497 vs. 7.533). For NCF-Platform-Centric compared to NCF, we see significant improvements in popularity (26.425 vs. 25.151) and hotel stars (3.764 vs. 3.740), while changes in helpfulness (1.377 vs. 1.379) and regional spread (7.533 vs. 7.542) are small and not significant. These patterns reveal a modest contrast with the RQ2 results: whereas user-centric

GenAI reviews approached human-authored content on helpfulness, platform-centric reviews do not. This highlights that while models trained on platform-generated reviews provide meaningful uplift relative to the baseline, they fall short of user-centric models in surfacing items linked to reviews that users perceive as more useful.

| Model | Avg Popularity@10 | Avg Helpfulness@10 | Avg Regional Spread@10 | Avg Hotel Stars@10 |
|---|---|---|---|---|
| Table 8: Business metrics comparison between original human reviews, platform-centric GenAI reviews with neutral tone and the NCF baseline. | | | | |
| NCF | 25.151 | 1.379 | 7.542 | 3.740 |
| NCF-Human | **27.632*** | **1.405** | **7.497**** | **3.813*** |
| NCF-Platform-Centric | 26.425 | 1.377 | 7.533 | 3.764 |

Note: * p ≤ 0.1, ** p ≤ 0.05, *** p ≤ 0.001 for the comparison between the NCF-Human and NCF-Platform-Centric

Taken together, the results for RQ3.a mirror the overall pattern observed in RQ2: human-authored reviews remain the gold standard for both performance and business outcomes, with statistically significant advantages across nearly all measures. Platform-centric reviews, however, still deliver a clear and meaningful uplift compared to the baseline. They significantly improve rating accuracy, popularity exposure, and hotel stars, and show positive though weaker gains on other measures. Two important implications follow. First, in situations of low user engagement, the platforms could generate draft reviews from interaction data and present them to users for optional editing, thus reducing user effort while increasing review volume and consequent RS performance. Second, and perhaps more importantly, platform-generated reviews can be leveraged behind the scenes: in cold-start or sparsity scenarios, they can be generated by the platform and fed into the training pipeline to improve performance. While not as effective as human-authored reviews, platform-centric reviews consistently outperform the no-review baseline, providing a valuable fallback when authentic content is limited.

### 6.3.2 RQ3.b: Implications of Platform-Centric Reviews under Cross Train–Test Conditions

In this experiment, we examined how platform-centric reviews affect model robustness under cross train–test conditions. Specifically, we tested whether models trained on one class of reviews could generalize when evaluated on another. At test time, each user's historical reviews were represented either in their original human-authored form or replaced with the corresponding platform-centric AI-generated version (as in RQ3.a). We considered two cross-source scenarios: (i) models trained on human reviews but tested on (neutral) platform-centric AI-generated reviews (Human→GenAI), and (ii) models trained on (neutral) platform-centric AI-generated reviews but tested on human reviews (GenAI→Human). This setup reflects realistic situations where a platform trains models on one review source but later operates in an environment dominated by another, thereby exposing the risks of distributional mismatch. We compared the performance of these models to that of models for which the training and test sets were of the same origin (Human→Human or GenAI→GenAI). As shown in Table 9, regardless of the composition of test set data, models trained on authentic human-authored reviews consistently achieved the strongest performance across all rating and ranking metrics, significantly outperforming models trained on platform-centric GenAI reviews. Specifically, for the test set comprising human-generated reviews, the Human→Human model outperformed the GenAI→Human model. Perhaps more notably, the human-trained model outperformed the GenAI-trained model even when the test environment was composed of platform-centric AI-generated reviews. Specifically, compared with the GenAI→GenAI model, the Human→GenAI model reduced RMSE from 1.052 to 1.011 (a 4% improvement), lowered MAE from 0.821 to 0.781, and increased MRR@10 from 0.063 to 0.077 (a 22% improvement) and NDCG@10 from 0.099 to 0.112 (a 13% improvement). All improvements were statistically significant, underscoring that human-authored content provides a more robust foundation for training recommendation models.

| Table 9: Rating and ranking performance comparison under cross train–test settings. | | | | |
|---|---|---|---|---|
| **Trained With→ Tested With** | **Rating Metrics** | | **Ranking Metrics** | |
| | **RMSE** | **MAE** | **MRR@10** | **NDCG@10** |
| (a) Human reviews → GenAI Reviews | **1.011***** | **0.781***** | **0.077***** | **0.112**** |
| (b) GenAI Reviews → GenAI Reviews | 1.052 | 0.821 | 0.063 | 0.099 |
| (c) Human Reviews→ Human Reviews | **1.014***** | **0.781***** | **0.078***** | **0.114**** |
| (d) GenAI Reviews → Human Reviews | 1.041 | 0.810 | 0.062 | 0.098 |

Note: * $p \leq 0.1$, ** $p \leq 0.05$, *** $p \leq 0.001$ for the comparison between (a)-(b) and (c)-(d).

We observed a similar pattern when evaluating RS performance with regard to business outcomes (Table 10). Specifically, when the test set comprised GenAI reviews, we observed that the Human→GenAI model outperformed the GenAI→GenAI model, surfacing higher-quality hotels (3.81 vs. 3.76), greater popularity exposure (27.5 vs. 26.4), and higher helpfulness (1.40 vs. 1.38). When the test set comprised human-authored reviews, the Human→Human model also achieved significantly better outcomes compared with the GenAI→Human model. For example, the Human→Human model achieved average hotel quality of 3.81 (vs. 3.77), and popularity exposure of 27.6 (vs. 26.40).

Together, the results of these analyses highlight the robustness of human-generated reviews as a source of training data, regardless of the composition of the deployment data (platform-centric GenAI or human-generated). In contrast, models trained on AI-generated reviews do not generalize as well to human-generated content. For platforms experiencing a shift toward AI-driven content, these findings suggest that the most effective strategy is to anchor model training in authentic human reviews while avoiding the use of GenAI reviews as training input. Nevertheless, our results also suggest that, when human-generated training data are unavailable (e.g., in cold-start scenarios), platform-centric reviews can serve as a useful source of training data. Specifically, training RS models on platform-centric reviews and deploying them on human-authored data still yields significantly better performance than relying on no textual information at all, as the NCF in Tables 7 and 8 performs significantly worse across all metrics.

| Trained With→ Tested With | Avg Popularity@10 | Avg Helpfulness@10 | Avg Regional Spread@10 | Avg Hotel Stars@10 |
|---|---|---|---|---|
| (a) Human reviews → GenAI Reviews | **27.51***** | **1.401***** | **7.491*** | **3.811***** |
| (b) GenAI Reviews → GenAI Reviews | 26.42 | 1.377 | 7.533 | 3.764 |
| (c) Human Reviews→ Human Reviews | **27.63***** | **1.405***** | **7.497*** | **3.813***** |
| (d) GenAI Reviews → Human Reviews | 26.43 | 1.376 | 7.538 | 3.768 |

Table 10: Business metrics comparison under cross train-test scenarios.

Note: * $p \leq 0.1$, ** $p \leq 0.05$, *** $p \leq 0.001$ for the comparison between (a)-(b) and (c)-(d).

### 6.3.3  RQ3.c: Implications of Tone-Based Framing Strategies for RSes

To assess whether stylistic framing enhances the effectiveness of platform-centric AI-generated reviews, we extended the platform-centric analysis introduced in RQ3.a. In this subsection, we evaluate three additional models that share the same architecture and training procedure as the NCF-Platform-Centric model but differ in the tone of the reviews used for training. While this model was trained on reviews generated under a neutral GenAI prompt, the alternative variants employ tone-conditioned prompts adopting Encouraging, Constructive, or Critical framings (see Table 2). This design isolates the effect of tone in the review-generation process, allowing us to examine whether evaluative framing can improve RS performance and downstream business outcomes. We benchmark these tone-conditioned models against both the neutral NCF-Platform-Centric model and the NCF-Human model, which represents authentic user-authored reviews.

As illustrated in Figure 6, tone framing, regardless of the tone itself, yields consistent gains across all rating metrics. RMSE improves from 1.052 under the neutral baseline to 1.030, 1.037, and 1.038 for the Encouraging, Constructive, and Critical models, respectively, all statistically significant reductions. MAE follows the same pattern, declining from 0.821 (for the neutral platform-centric baseline) to 0.788, 0.802, and 0.803, with the improvements for Encouraging and Constructive reaching significance. Results for the ranking metrics point in the same direction, though the effects are somewhat smaller. MRR@10 rises from 0.063 in the neutral model to 0.071 for both Encouraging and Critical framings and 0.068 for Constructive, with the first two reaching statistical significance. NDCG@10 increases modestly from 0.099 to 0.103–0.106 but does not achieve significance.

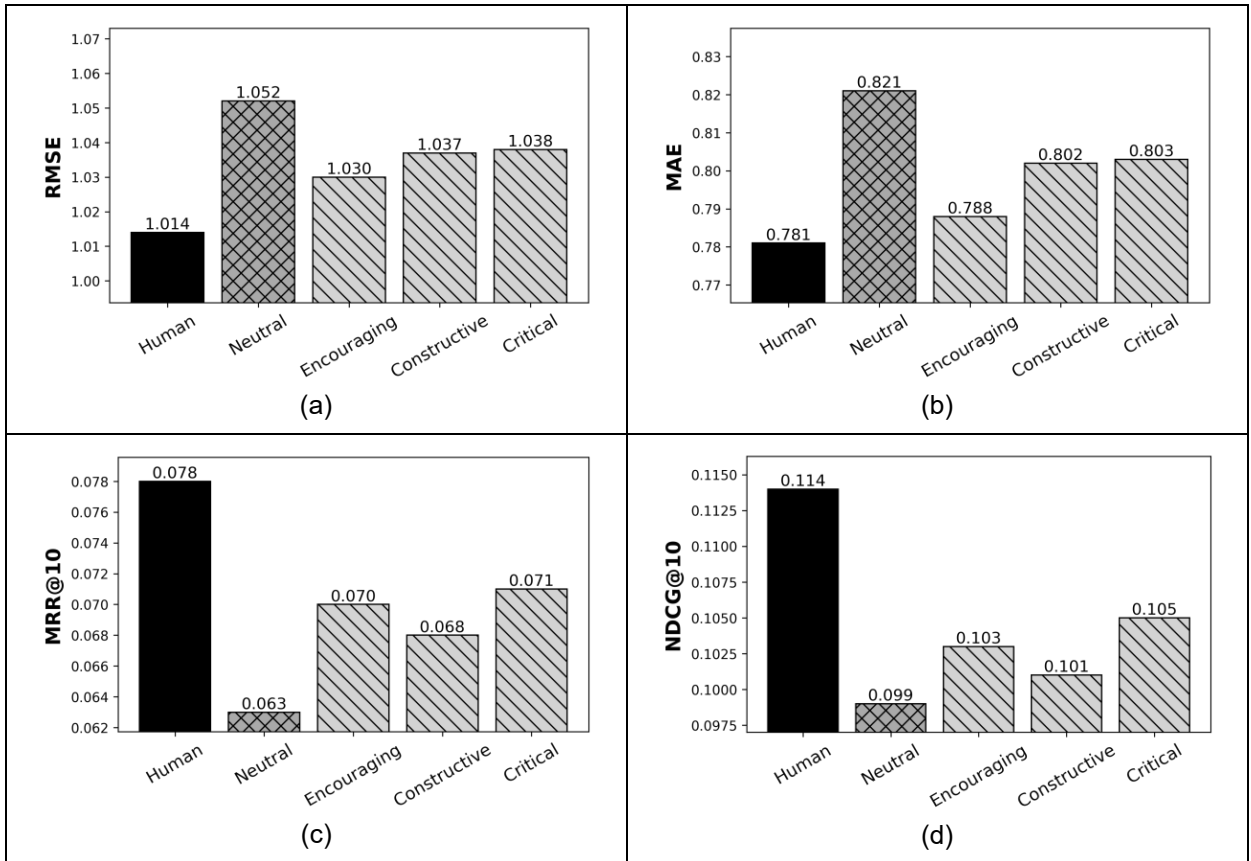

Figure 6: Comparison of platform-generated review tones across ranking and rating metrics.

Taken together, these results suggest that stylistic framing (relative to neutral text) helps infuse AI-generated reviews with clearer evaluative signals, which improve model learning and enhance rating prediction accuracy, though ranking precision gains are more moderate. Nevertheless, when compared to NCF-Human, tone-based platform-centric models still fall short across all performance measures (e.g., RMSE = 1.014, MRR@10 = 0.078).

Figure 7 reports the business outcomes, which mirror these performance trends. All tone-framed models significantly outperform the neutral baseline along the measures of popularity, helpfulness, and hotel quality. For example, average popularity increases from 26.425 to 26.947–27.374, hotel quality rises from 3.764 to 3.777–3.803, and helpfulness scores improve from 1.377 to 1.39–1.41. These shifts indicate that even minimal tone conditioning enhances how the recommender surfaces informative, well-regarded items, thereby aligning recommendations more closely with business value. Regional spread shows mixed fluctuations without a consistent directional trend. Interestingly, when we compare the performance of the tone-based models to the performance of NCF-Human for these focal business metrics, the gaps observed in earlier analyses (comparing the neutral NCF-Platform-Centric model to NCF-Human) for helpfulness and hotel quality become statistically insignificant. This narrowing, coupled with the consistent and significant improvements over the neutral baseline in our evaluations of ranking and rating performance, demonstrates the promise of tone framing as a low-cost, high-impact enhancement to AI-generated reviews. By introducing minimal stylistic variation, these models already move closer to human-authored performance and can meaningfully enhance the quality and business impact of an RS trained on AI-generated reviews.

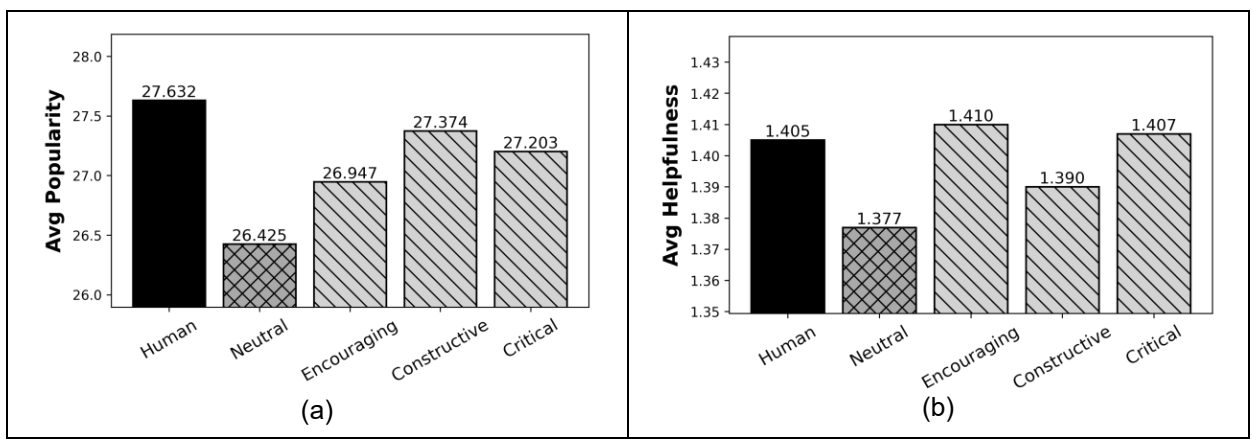

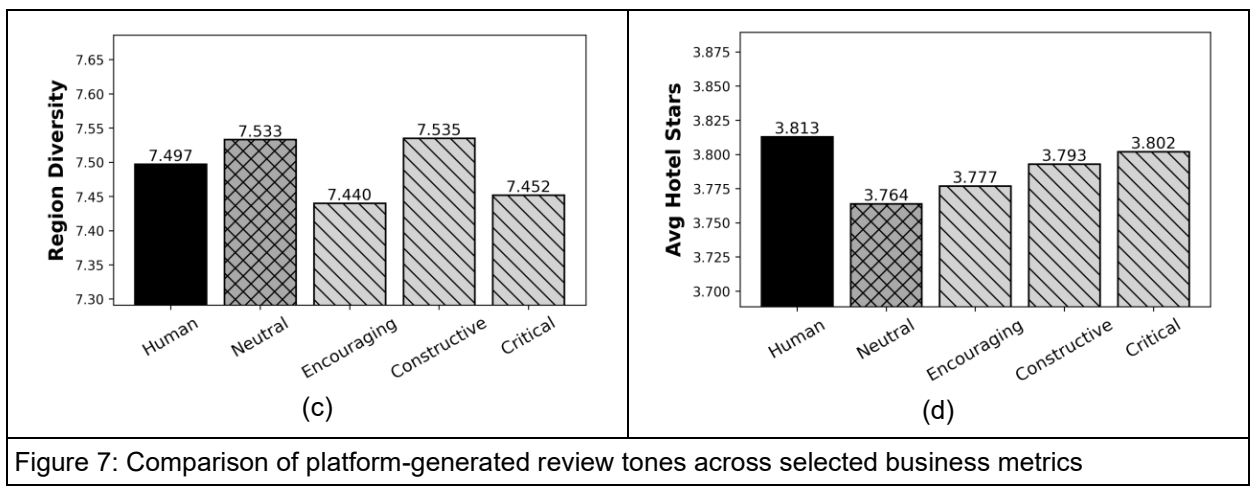

Figure 7: Comparison of platform-generated review tones across selected business metrics

## 7 DISCUSSION

In this paper, we have provided a systematic investigation of how LLM-generated reviews impact recommender system performance along various rating/ranking metrics and business outcomes. We examined this phenomenon through two lenses: (1) user-centric scenarios, where individuals employ AI tools to refine their own reviews, and (2) platform-centric scenarios, where platforms generate content from explicit interactions (e.g., star ratings), for example, to address cold-start challenges where textual reviews are missing, or as a means of compensating for users' reluctance to write their own reviews. Our findings reveal that the integration of AI-generated reviews alters the information ecosystem that RSes rely upon, with measurable consequences for both algorithmic performance and strategic platform objectives.

First, we textually analyzed corpora of human-generated and AI-generated reviews that were produced to simulate either user-centric or platform-centric scenarios (RQ1). These analyses revealed systematic differences between human and AI-generated content, consistent with the findings of previous research. Specifically, for both user- and platform-centric scenarios, AI-generated reviews demonstrated higher internal similarity, reduced lexical diversity, narrower emotional expression, and sentiment shifts in comparison with human-generated text. In RQ2 (focusing on user-centric scenarios) and RQ3.a (neutral platform-centric scenarios), we demonstrated that these textual differences translate into reduced recommendation performance. While models trained and deployed on either human- or AI-generated reviews outperformed a traditional collaborative filtering model that did not integrate textual review data (the NCF model), the magnitude of improvement differed substantially. For example, in the case of user-centric GenAI reviews in RQ2, an RS trained and tested on human reviews reduced the rating prediction error (RMSE) by approximately 12.1% compared to no reviews, while RSes trained on user-centric or on platform-centric AI content achieved reductions of only 10.8% and 8.8%, respectively. These differences also extended to business value metrics: Compared with either user-centric or platform-centric GenAI reviews, human-authored reviews enabled the system to surface higher-quality and more popular hotels, as well as hotels that were associated with more helpful reviews. Notably, however, compared with the no-review baseline, GenAI reviews still delivered consistent gains across business value metrics, with user-centric reviews providing particularly meaningful uplift, in addition to broadening geographic coverage. Together, these findings confirm that the origin and framing of review content shape not only predictive accuracy but also downstream business outcomes.

In RQ3.b we delved further into the platform-centric scenario by examining a cross train-test setting, in which RSes were trained on human-authored reviews and tested on platform-centric GenAI reviews, or vice versa. Our experiments revealed a notable asymmetry in which systems trained on human reviews obtained better performance when tested on either human or AI-generated content, whereas systems that trained on AI-generated reviews exhibited poorer performance than human-trained models across both test conditions. These findings suggest that human-authored reviews provide superior model training performance, enabling recommendation models to generalize more effectively regardless of the content origin on which they are deployed. That is, when RSes are trained on human reviews, injecting AI-generated reviews in the deployment phase will still provide stable performance. By contrast, training on AI-generated reviews produces less resilient systems, degrading performance across deployment contexts. From a business perspective, human-trained models consistently surface higher-quality hotels and maintain these standards even when tested on AI content. In contrast, models trained on AI-generated reviews tend to recommend lower-quality items, underscoring how the training data source fundamentally shapes a system's ability to identify and prioritize quality signals.

These findings highlight the importance of detecting and carefully managing the injection of generative AI content into the training process of recommendation systems. Though platform-centric generation of reviews from interaction data is likely to prove beneficial in situations where no textual review data is available, it is preferable to train RSes on human-generated reviews when possible.

Finally, in RQ3.c, we used GenAI tools to generate synthetic reviews strategically designed with distinct framings, encouraging, constructive, or critical, that adopt particular tones. This setup allowed us to examine how semantic and lexical shifts in review text influence RS performance. All three tone variants significantly outperformed the neutral baseline, demonstrating that such tonal diversity consistently improves both recommendation accuracy and business outcomes, including indicators such as popularity exposure, hotel quality, and helpfulness. These findings further suggest that platforms can actively control the generative AI data they inject into the training process, strategically designing review tone (e.g., emphasizing encouraging or constructive feedback) to enhance overall system effectiveness. Importantly, tone-conditioned synthetic reviews move closer to the signal provided by human-authored input than neutral AI text, showing that even modest stylistic shifts help bridge the gap between artificial and authentic reviews, with several business metrics reaching parity. These results demonstrate that prompt-based framing is more than a technical optimization; it represents a strategic lever for platforms to improve the effectiveness of their personalization methods using AI-generated reviews. By deliberately shaping synthetic content through tone, platforms can consistently boost both recommendation accuracy and business outcomes without sacrificing user privacy or exposing the generative content itself, leveraging it solely within the training of the RS model.

## 7.1  Managerial Implications

Our results highlight several considerations for managers of digital platforms integrating GenAI review data into RSes.

**Leverage AI reviews but cautiously**: While AI-generated reviews outperform a no-review baseline, both in terms of recommender performance and business metrics, they are still outperformed by RSes trained on human-authored content. Managers should therefore treat AI-generated reviews within the RS pipeline as a complementary tool for mitigating cold-start or sparse-data challenges, rather than as a full substitute for authentic human input.

**Gen AI can pollute/bias RS training, so training should be anchored in human content**: In our analyses, models trained on human reviews generalized robustly to both human- and AI-generated review data in the test sets, whereas models trained on AI reviews achieved poorer performance across test sets comprising either human- or AI-generated reviews. These findings underscore the strategic value of maintaining human-authored reviews as the foundation for recommender training.

**Detect and align GenAI content**: Given that GenAI review content can alter how RSes learn and perform, platforms should develop tools that can detect AI-influenced content (e.g., using detection methods such as those described by Santos and Antonio (2025)). Behind the scenes, within the RS training and model-building processes, these tools can inform strategies such as filtering or re-aligning AI-augmented reviews to better reflect human-authored tone and sentiment. Such measures enable platforms to integrate AI-enhanced content more effectively, ensuring that it supports, rather than undermines, RS performance.

**Inject strategic tones optimized for decision-making in GenAI reviews**: Our findings reveal that platform-centric review generation can be strategically guided to improve recommendation quality. Specifically, when reviews are optimized using intentional stylistic tones, such as encouraging, constructive, or critical, the resulting content improves RS performance relative to neutral, untuned outputs. These types of reviews narrowed the performance gap between AI-generated and human-authored reviews, suggesting that the way platforms instruct AI to frame and stylize review content can materially influence recommendation outcomes. While human-authored reviews remain the gold standard, platforms can strategically guide GenAI content to complement human input. Tone-aware prompting thus emerges as a simple yet effective strategy for enhancing personalization and the overall RS performance.

## 7.2   Limitations and Future Research

This study is not without limitations, which also point to avenues for future work. Our analysis focused on the travel domain, which encompasses a major and well-studied area of online consumer decision-making. In this domain, reviews play a central role in shaping consumers' decisions and platform credibility (Santos and Antonio, 2025). While this makes it a highly relevant and informative context for our study, it may not fully capture the dynamics of other verticals such as retail, entertainment, or social media. These domains differ in both the nature of their content and the behavioral patterns of their users. Future research should therefore examine whether the patterns observed in this study generalize across different types of platforms, where review styles, content characteristics, and user behaviors may vary.

Another limitation lies in our choice of RS architecture. We employed Neural Collaborative Filtering (NCF), a widely used and industry-validated state-of-the-art neural baseline that has continued to achieve strong accuracy in recent benchmark studies (Do and Nguyen, 2022; Wu et al., 2022). While we focus on NCF as the core algorithm for our analysis, other recommendation models, such as attention-based architectures, graph neural networks, or sequence-aware models, may interact with AI-generated reviews in different ways. Such work could open opportunities to uncover architecture-specific vulnerabilities or benefits when training RSes on GenAI content.

In addition, our exploration of strategic review generation was guided by prior IS research emphasizing the importance of emotional and tonal cues in RSes (e.g., Yu et al., 2023; Kim and Hong, 2024; Abakarim et al., 2025). We examined three intentional stylistic tones for generating textual reviews (constructive, encouraging, and critical) and found that these tone-based instructions improved recommendation performance relative to neutral, untuned outputs. While our focus in this study was not on optimal review generation, we believe that future research could extend this approach by investigating additional sophisticated forms of textual guidance that can further narrow the gap between AI-generated and human-authored reviews.

Finally, our focus in this study was not on detecting GenAI content per se. Rather, we showed that the inclusion of AI-generated material has a measurable effect on system performance. Future research could simultaneously address the problems of both detecting and mitigating the effects of AI-generated content. As AI-produced content becomes increasingly widespread and commonplace, it will no longer be feasible to keep training data purely human-authored. Future work should therefore address how platforms can adapt, by detecting and flagging AI-generated content where relevant, while also developing strategies to integrate or counterbalance it, rather than assuming a clean separation between human and AI inputs.